\documentclass[preprint, 12pt]{elsarticle}

\usepackage{amssymb}
\usepackage{algorithm} 
\usepackage{algpseudocode}
\usepackage{amsmath}
\usepackage{makecell}
\usepackage{xcolor}
\usepackage{float}

\usepackage{pdfpages} 
\usepackage{afterpage}
\usepackage{lipsum} 
\usepackage{fancyhdr} 
\usepackage{hyperref}
\usepackage{glossaries-extra}
\makeatletter
\newcommand*{\glsplainhyperlink}[2]{%
  \colorlet{currenttext}{.}
  \colorlet{currentlink}{\@linkcolor}
  \hypersetup{linkcolor=currenttext}
  \hyperlink{#1}{#2}%
  \hypersetup{linkcolor=currentlink}
}
\let\@glslink\glsplainhyperlink
\makeatother

\setabbreviationstyle[acronym]{long-short}
\glssetcategoryattribute{acronym}{nohyperfirst}{true} 

\newacronym{tte}{TTE}{transthoracic echocardiography}
\newacronym{tee}{TEE}{transesophageal echocardiography}
\newacronym{us}{US}{ultrasound}
\newacronym{lax}{LAX}{long-axis}

\newacronym[plural=SHDs, firstplural=Structural heart diseases]{shd}{SHD}{structural heart disease}
\newacronym[plural=CVDs, firstplural=Cardiovascular diseases]{cvd}{CVD}{cardiovascular disease}
\newacronym[plural=MVs, firstplural=mitral valves]{mv}{MV}{mitral valve}
\newacronym{la}{LA}{left atrium}
\newacronym{lv}{LV}{left ventricle}
\newacronym{mr}{MR}{mitral regurgitation}
\newacronym{ma}{MA}{mitral annulus}
\newacronym{mvp}{MVP}{mitral valve prolapse}
\newacronym{teer}{TEER}{transcatheter edge-to-edge repair}
\newacronym{sh}{SH}{saddle horn}
\newacronym{cc}{CC}{commissure-commissure}
\newacronym{pam}{PAM}{posterior annular midpoint}
\newacronym{es}{ES}{end sytole}
\newacronym{ed}{ED}{end diastole}
\newacronym{prelu}{PReLU}{parametric rectifying linear unit}
\newacronym[plural=ReLUs, firstplural=rectifying linear units]{relu}{ReLU}{rectifying linear unit}

\newacronym{ai}{AI}{artificial intelligence}
\newacronym{ar}{AR}{augmented reality}
\newacronym{ml}{ML}{machine learning}
\newacronym{dl}{DL}{deep learning}
\newacronym[plural=NNs, firstplural=Neural networks]{nn}{NN}{neural network}
\newacronym[plural=CNNs, firstplural=Convolutional neural networks]{cnn}{CNN}{convolutional neural network}
\newacronym{vit}{ViT}{vision transformer}
\newacronym{tl}{TL}{transfer learning}
\newacronym{ssl}{SSL}{semi-supervised learning}
\newacronym{sda}{SDA}{strong data augmentation}
\newacronym{gan}{GAN}{generative advesarial Network}
\newacronym[plural=GTs, firstplural=ground truths]{gt}{GT}{ground truth}
\newacronym{cad}{CAD}{computer-aided design}
\newacronym{gnn}{GNN}{graph neural network}
\newacronym{mlp}{MLP}{multi-layer perceptron}
\newacronym{lb}{LB}{Laplace-Beltrami}
\newacronym{pca}{PCA}{principal component analysis}
\newacronym{icp}{ICP}{iterative closest point}

\newacronym[
    first={average surface distance (ASD, in mm)},
    long={average surface distance},
    short={ASD}
]{asd}{ASD}{average surface distance}
\newacronym[first={Hausdorff Distance 95\% (95\% HD, in mm)}, \glslongpluralkey={Hausdorff Distances 95\%}, \glsshortpluralkey={HDFs 95\%}]{hdf95}{95\% HD}{95\% Hausdorff Distance}
\newacronym{sdf}{SDF}{Signed Distance Function}
\newacronym{tcs}{TCS}{temporal consistency score}
\newacronym{aad}{AAD}{average absolute deviation}

\newacronym{ge}{GE HealthCare}{General Electric HealthCare}
\newacronym{ntnu}{NTNU}{Norwegian University of Science and Technology's}

\journal{ToBeDefined}

\begin{document}

\begin{frontmatter}



\title{MitraClip Device Automated Localization in 3D Transesophageal Echocardiography via Deep Learning}


\author[Poli]{Riccardo Munaf\`{o}}
\ead{E-mail address: riccardo.munafo@polimi.it, Postal address: Via Giuseppe Ponzio, 34/5, 20133 Milano MI}
\author[Poli,aumc,uva]{Simone Saitta}
\author[Simulands]{Luca Vicentini}
\author[Poli]{Davide Tondi}
\author[Poli]{Veronica Ruozzi}
\author[PSD,Poli]{Francesco Sturla}
\author[OSR]{Giacomo Ingallina}
\author[Simulands]{Andrea Guidotti}
\author[OSR,Uni-OSR]{Eustachio Agricola}
\author[Poli]{Emiliano Votta}

\affiliation[Poli]{organization={Department of Electronics, Information and Bioengineering, Politecnico di Milano},
            city={Milan},
            country={Italy}}
\affiliation[aumc]{organization={Department of Biomedical Engineering and Physics, Amsterdam UMC},
            city={Amsterdam},
            country={The Netherlands}}
\affiliation[uva]{organization={Informatics Institute, University of Amsterdam},
            city={Amsterdam},
            country={The Netherlands}}
\affiliation[Simulands]{organization={Simulands},
            city={Zurich},
            country={Switzerland}}
\affiliation[PSD]{organization={3D and Computer Simulation Laboratory},
            city={Milan},
            country={Italy}}
\affiliation[OSR]{organization={Unit of Cardiovascular Imaging, IRCCS San Raffaele Hospital},
            city={Milan},
            country={Italy}}
\affiliation[Uni-OSR]{organization={Vita-Salute San Raffaele University},
            city={Milan},
            country={Italy}}

\begin{abstract}
The MitraClip is the most widely percutaneous treatment for mitral regurgitation, typically performed under the real-time guidance of 3D transesophagel echocardiography (TEE). However, artifacts and low image contrast in echocardiography hinder accurate clip visualization. This study presents an automated pipeline for clip detection from 3D TEE images. An Attention UNet was employed to segment the device, while a DenseNet classifier predicted its configuration among ten possible states, ranging from fully closed to fully open. Based on the predicted configuration, a template model derived from computer-aided design (CAD) was automatically registered to refine the segmentation and enable quantitative characterization of the device. The pipeline was trained and validated on 196 3D TEE images acquired using a heart simulator, with ground-truth annotations refined through CAD-based templates. The Attention UNet achieved an average surface distance of 0.76 mm and 95\% Hausdorff distance of 2.44 mm for segmentation, while the DenseNet achieved an average weighted F1-score of 0.75 for classification. Post-refinement, segmentation accuracy improved, with average surface distance and 95\% Hausdorff distance reduced to 0.75 mm and 2.05 mm, respectively. This pipeline enhanced clip visualization, providing fast and accurate detection with quantitative feedback, potentially improving procedural efficiency and reducing adverse outcomes.

\end{abstract}




\end{frontmatter}


\section{Introduction}
\label{sec:Introduction}


\Gls{teer} is the most widespread percutaneous treatment for \gls{mr} \cite{Vahanian_2021}. It offers a safe and effective alternative for patients with contraindications for surgery or those at high operative risk \cite{Feldman_2015, Buzzati_2019}. The MitraClip (Abbott Laboratories, California, USA) is a catheter-based technology for \gls{mv} \gls{teer}, designed to treat \gls{mr} by clipping together the mitral leaflets. The procedure involves accessing the \gls{la} through the interatrial septum with a steerable sheath. A delivery catheter is then advanced in the sheath and steered through the \gls{la} to reach the \gls{mv} region where lack of coaptation is observed. Upon reaching this target region, a clip, located on the tip of the delivery catheter, is actuated to grasp the \gls{mv} leaflets and it is finally deployed. To achieve optimal grasping, the clip should be oriented perpendicularly to the \gls{mv}, with its arms locally orthogonal to the coaptation line to be restored. \\
These steps, from transseptal puncture through the steering of the delivery catheter toward \gls{mv} and to clip positioning, are performed under the guidance of 2D and 3D \gls{tee} \cite{Sherif_2017}. 3D \gls{tee} provides an \textit{en face view} of the \gls{mv} and of the approaching clip, allowing clinicians to determine when the clip is adequately positioned and oriented with respect to the target region of the \gls{mv} leaflets \cite{Sherif_2017}. 3D \gls{tee} images can be non-trivial to interpret, since they are displayed on 2D monitors, with a loss of depth perception. Thus, clinicians often rely on 2D views for clip positioning, which requires adjusting the \gls{tee} transducer to align the clip in 2D \gls{tee} views or extracting specific slices from the 3D volume for better visualization on the fly \cite{Nyman_2018}. However, both actions are time-consuming and operator-dependent, and require specific expertise. Moreover, the presence of the catheter and of the clip generates artifacts that hamper the assessment of clip configuration, in particular when it is close to native tissues.  
To address these challenges, an automated, real-time clip detection system for 3D \gls{tee} could provide significant benefits. Such a system would enable the automatic extraction of relevant 2D views encompassing the device, and, when integrated with automated \gls{mv} analysis \cite{Munafo_2024a}, facilitate precise quantification of the spatial relationship between the clip and its target region. This would enhance procedural guidance by providing standardized, operator-independent visualization.\\

\subsection{Related works}
\label{subsec:Related Works}

Despite the clinical importance of intraoperative guidance, no prior study has tackled the automated segmentation of \gls{tee} for characterizing the position and configuration of clip devices for \gls{teer}. Published studies have primarily focused on catheter localization in \gls{tee} images, leveraging detection or segmentation techniques \cite{Yang_2023}. Detection methods aim to identify the general location or shape of the catheter by using representations such as bounding box (a rectangular region enclosing the catheter), axis (a central line running along the catheter's length), or skeleton (a simplified line-based representation capturing the catheter's structure). These methods typically rely on carefully designed filters or instrument templates to extract catheter-related information from the image \cite{Yang_2023}. On the other hand, segmentation involves classifying each voxel associated with the catheter to reconstruct the latter within the image. \Glspl{cnn} are the state-of-the-art tools for automatic segmentation of 3D echocardiography images, and recent studies have explored their application to cardiac catheter localization. Yang et al. applied a 2D \gls{cnn} architecture for catheter segmentation in 3D \gls{tee} by stacking together adjacent 2D axial slices \cite{Yang_2019}. To mitigate the possible loss of 3D semantic information after slicing, they subsequently introduced image patch extraction and multi-planar slicing \cite{Yang_2019b}. Although the patch-based approach benefits from speed and low GPU memory usage, it may limit the network's ability to capture the full contextual information of the image. To address this, Yang et al. adopted a hybrid loss function that integrates voxel-wise loss with contextual loss for 3D UNet training, encouraging the network to learn a better contextual representation \cite{Yang_2019c, Yang_2021}. Alternatively, full 3D image information has been leveraged by combining 3D encoder and projection layers for features dimension reduction along axial and lateral dimensions, achieving efficient catheter segmentation in 3D \gls{tee} \cite{Yang_2020}. Despite these advancements, automatic catheter segmentation still requires additional complex post-processing to extract meaningful clinical information, such as catheter position and orientation. Furthermore, \gls{cnn} methods for semantic segmentation rely on accurate \gls{gt} annotations, which are difficult to obtain due to the noise and artifacts present in \gls{tee} images when visualizing metallic catheters. This frequently leads to over-segmentation of the catheter, producing numerous false positives that negatively affect the localization \cite{Mastmeyer_2017}. Consequently, these approaches have been applied to detect intracardiac catheters with simple shapes, such as ablation catheters or guide wires that are typically characterized by a straight configuration during the procedure. \\

\subsection{Main contribution}
\label{subsec:Contibutions}

In this work, we present the first automated approach for detecting, localizing, and characterizing the clip on the tip of the MitraClip delivery catheter. For clarity, we will refer to the MitraClip device, specifically the component involved in the grasping of \gls{mv} leaflets during the procedure, simply as the clip. Unlike previous efforts that focused solely on catheter segmentation, our method is designed to accurately extract the clip’s spatial features, which are essential for intraoperative guidance during \gls{teer}. We introduced a novel deep learning-based framework for clip detection in volumetric echocardiography data. Beyond simple segmentation, our approach enabled a comprehensive characterization of clip orientation and positioning, providing key parameters to support real-time surgical decision-making. Our method reduces the reliance on manual slice selection and expert interpretation, allowing for standardized and efficient intraoperative visualization. By addressing these challenges, our approach has the potential to enhance procedural accuracy, reduce operative time, and improve overall patient outcomes in \gls{teer} interventions.

\section{Methods}
\label{sec: Methods}

\subsection{Dataset Collection}
\label{subsec: Dataset Collection}
To implement and validate our proposed clip detection method, we collected 196 4D \gls{tee} recordings in a heart simulator designed to replicate transcatheter \gls{mv} repair procedures \textit{in vitro}. The simulator included anatomically realistic phantoms of:

\renewcommand\labelitemi{$\vcenter{\hbox{\tiny$\bullet$}}$}

\begin{figure}[ht]
    \centering
    \centerline{\includegraphics[width=\textwidth]{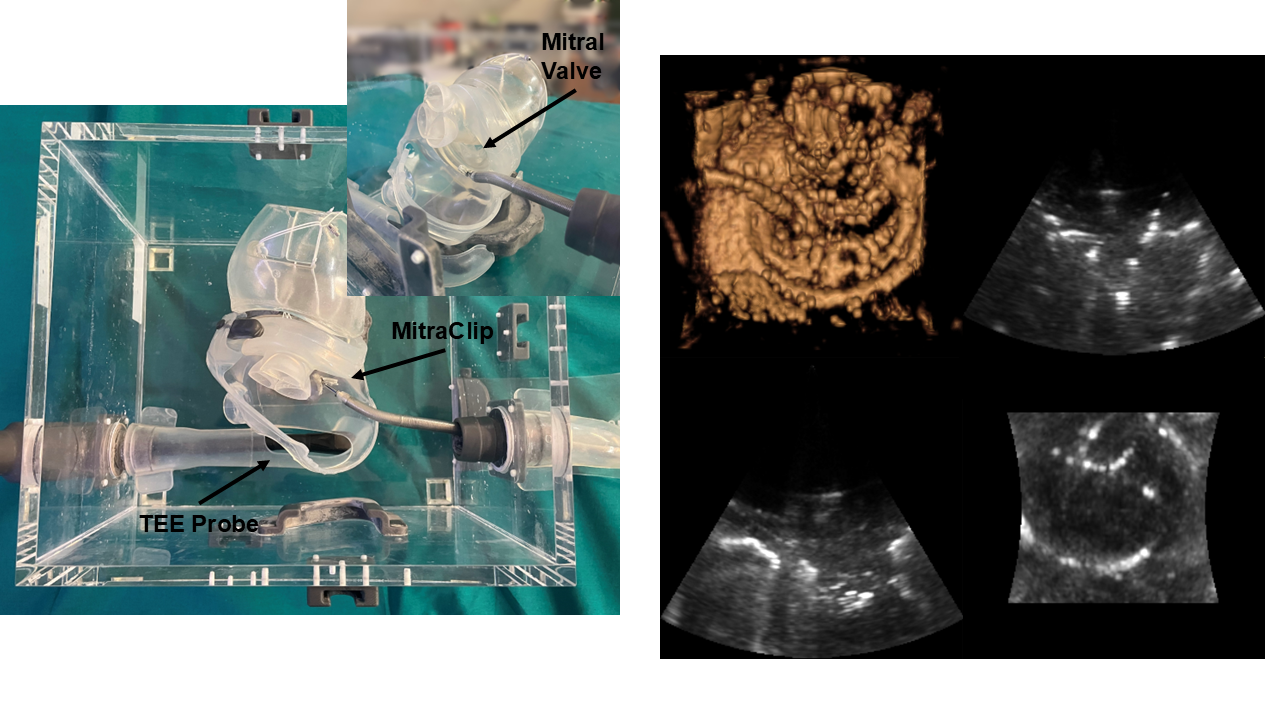}}
    \caption{Left: \textit{in-vitro} setup for \gls{tee} acquisitions in \gls{teer} procedures. Black arrows indicate the \gls{tee} probe, the MitraClip and the \gls{mv}. Right: example of \textit{In-vitro} 3D \gls{tee} acquisition with MitraClip catheter. }
    \label{fig: experimental set-up}
\end{figure}

\begin{itemize}
    \item esophagus
    \item access veins, i.e., femoral, iliac, and hepatic veins as well as the inferior vena cava connected to the right atrium
    \item heart, including the relevant intracardiac structures, i.e., interatrial septum, left atrium, and \gls{mv} with moving leaflets
\end{itemize}
All phantoms were immersed in demineralized water to enable \gls{tee} imaging. 
The setup was complemented by a MitraClip system (Abbot, G4 version) with a XTW clip and by a GE Vivid E95 scanner (GE Vingmed Ultrasound, Horten, Norway) with 4Vc-D \gls{tee} probe.
At the beginning of experiments, the \gls{tee} probe was advanced next to the heart phantom through the esophagus phantom. It was then kept the probe at midesophageal level, ensuring procedural views of both \gls{mv} and the MitraClip catheter in 2D and 3D modes. 
The MitraClip catheter was inserted in the phantom of the right femoral vein and driven through the interatrial septum and into the left heart (Figure \ref{fig: experimental set-up}). We followed the procedural guidelines for \gls{teer} with the MitraClip system \cite{Sherif_2017, Nyman_2018} for positioning the clip above the \gls{mv}. The steps for positioning, which are preparatory for the deployment of the clip in either the central, lateral, or medial portion of the \gls{mv}, included:
\begin{enumerate}
    \item Advancing the clip delivery system through the sheath into the \gls{la}.
    \item Steering the sheath catheter and simultaneously medially deflecting the clip delivery system while retracting the whole system to position the clip delivery system above the \gls{mv}. This step included gentle anterior rotation of the clip delivery catheter accompanied by simple lateral rotations.
    \item Adjusting the medial-lateral clip position and opening the clip.
\end{enumerate}

\Gls{tee} imaging was recorded at every stage of the procedure following the access of the delivery catheter into the \gls{la}. When the clip was close to the \gls{mv}, imaging was acquired for each combination of clip pose and opening angle. Of note, in every recording the \gls{mv} was positioned approximately at the center of the 3D volume. For each recording, a single frame corresponding to a 3D \gls{tee} image was extracted. 196 3D \gls{tee} images were obtained and exported in Cartesian format and sampled with a uniform spacing of 0.5 mm, resulting in a mean resolution of $199 \times 248 \times 248$.

\subsection{Dataset Annotation}
\label{subsec: Dataset Annotation}

\subsubsection{Clip Manual Annotation}
\label{subsubsec: Clip Raw Segmentation}

Two trained users segmented the MitraClip delivery catheter and clip, both assigned with the same label, using 3D Slicer \cite{Kikinis_2013}. Voxel thresholding was applied using the Otsu method \cite{Otsu_1979}, and the threshold intensity range was adjusted to isolate voxels corresponding to the MitraClip catheter and heart structures from the background. Using the 3D-rendered visualization of the segmentation, the users manually refined the segmentation by cutting away areas that were not considered part of the MitraClip catheter. 

\subsubsection{Design of MitraClip template models}
\label{subsubsec: Clip Template Models}

A real XTW clip, detached from its delivery catheter, was scanned using a 3D scanner (Artec Micro II, Artec 3D Technology, Senningerberg, Luxembourg) to digitize the object. The clip was scanned in three configurations: fully closed and open with opening angle ($\theta$) equal to 60° and 120°, respectively. Scanned models were exported in \textit{.stl} format and were used as references to sketch the template models of the clip through \gls{cad} modeling in Autodesk Fusion (Autodesk Inc., San Francisco, California, United States). We started designing the MitraClip model in closed configuration (Figure \ref{fig:gt refinement}, top row, right). The model was conceived as a two-part 3D object, and we exploited the longitudinal symmetry of the clip by creating only one half of it. The complete model was then obtained by mirroring the generated half. The first part represented the bottom portion of the clip, extending from the tip to the origin of the clip's arms begin (Figure \ref{fig:gt refinement}, top row, in red). The upper part consisted of the clip's arm (Figure \ref{fig:gt refinement}, top row, in green). The bottom part was created using the loft method by fitting three elliptical half sections sketched along the clip axial direction at 1 mm, 3 mm, and 6 mm from the tip. The dimensions of these sections were chosen to match the corresponding segments of the scanned model. Similarly, the upper part was created by fitting two elliptical half sections sketched at 6 mm and 18 mm from the clip’s tip and fitted using the loft method. Then, nine template models were automatically generated by rigidly tilting the arm. These were characterized by $\theta$ ranging from 10° up to 90° (10° step). To account for the actual deformation of the real object and ensure the accuracy of the templates, for the templates with $\theta$ equal to 60° and 120° the length and the cross-sectional dimensions of the arm were adjusted to match their values measured in the corresponding scanned models. These were linearly interpolated to obtain the length and the cross-sectional dimensions of the arm in the other open configurations, thus representing the gradual changes in clip morphology across its range of configurations. 

\subsubsection{Clip Annotation CAD-based Refinement}
\label{subsubsec: Clip Refined Ground Truth}

To obtain accurate \gls{gt} data, manual segmentations were automatically refined using the clip templates generated through \gls{cad} modeling. To this aim, given a manual segmentation mask, the following steps were implemented:
\begin{enumerate}
    \item the manual segmentation mask was converted into a triangulated surface \((\Omega_s)\) by marching cubes algorithm;
    \item the extremities of \((\Omega_s)\) were identified as the vertices associated with the minimum and maximum value of the first eigenvalue of the \gls{lb} operator \cite{Reuter_2006};
    \item the extremity closest to the center of the image, and hence to the \gls{mv}, was identified as the tip of the clip ($\textbf{C}_s$);
    \item the longitudinal and transversal axes ($\hat{\textbf{n}}_{s}$ and $\hat{\textbf{t}}_{s}$) of $\Omega_{s}$ were identified as the first and second principal direction, respectively, yielded by \gls{pca} on the 3D coordinates of the vertices of $\Omega_{s}$;
    \item each template ($T_i$), characterized by clip tip ($\textbf{C}_{ti}$), longitudinal axis ($\hat{\textbf{n}}_{ti}$), and transversal axis ($\hat{\textbf{t}}_{ti}$), was rigidly co-registered on the segmentation mask so to have $\textbf{C}_{ti}$ superimposed to $\textbf{C}_{s}$, and $\hat{\textbf{n}}_{ti}$ and $\hat{\textbf{t}}_{ti}$ aligned with $\hat{\textbf{n}}_{s}$ and $\hat{\textbf{t}}_{s}$, respectively;
    \item the Dice score between each co-registered template $T_i$ and the segmentation mask was computed;
    \item the co-registered template with maximal Dice score was used to refine the raw segmentation mask.
\end{enumerate}
 
\begin{figure}[ht]
    \centering
    \centerline{\includegraphics[width=\textwidth]{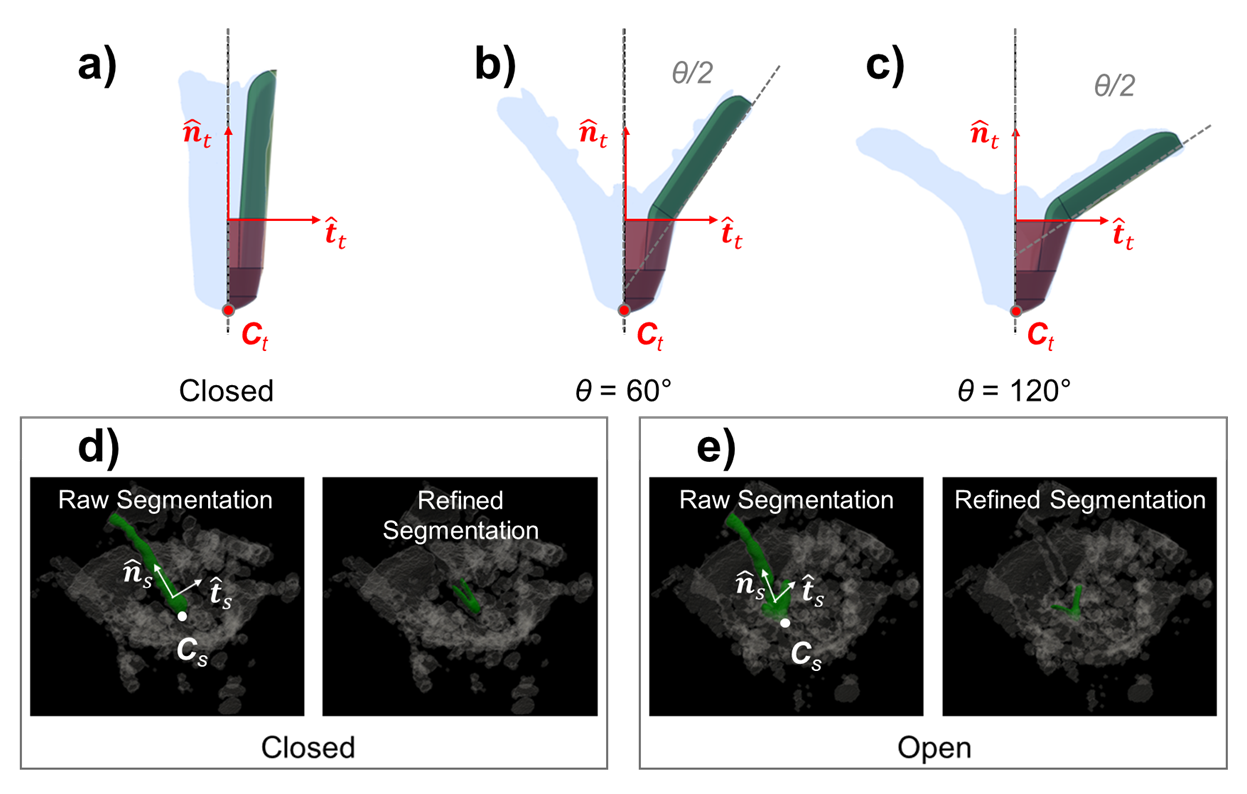}}
    \caption{a)-c): \gls{cad} models of the clip in three different configurations. Only one longitudinal half of the \gls{cad} model is shown with two different color for the bottom part (in red) and the upper part (in green). 3D representations of the scanned real clip are shown in transparency at the corresponding configurations. The clip tip (\(\textbf{C}_{t}\)), longitudinal axis (\(\hat{\textbf{n}}_{t}\)) and transversal axis (\(\hat{\textbf{t}}_{t}\)) are highlighted in red for each template.
    \\
    d)-e): Two examples of the clip \gls{gt} segmentations refinement using the template models in closed and open configurations. For each example, the raw segmentation mask is depicted on the left hand side, including its clip tip (\(\textbf{C}_{s}\)), longitudinal axis (\(\hat{\textbf{n}}_{s}\)) and transversal axis (\(\hat{\textbf{t}}_{s}\)), while the refined mask is shown on the right hand side.}
    \label{fig:gt refinement}
\end{figure}

Figure \ref{fig:gt refinement}.d and Figure \ref{fig:gt refinement}.e, show two examples of \gls{gt} segmentation before and after refinement through template registration for clips in closed and open configurations, respectively.

\subsection{Application Pipeline}
\label{subsec:Application Pipeline}

\begin{figure}[ht]
    \centering
    \centerline{\includegraphics[width=\textwidth]{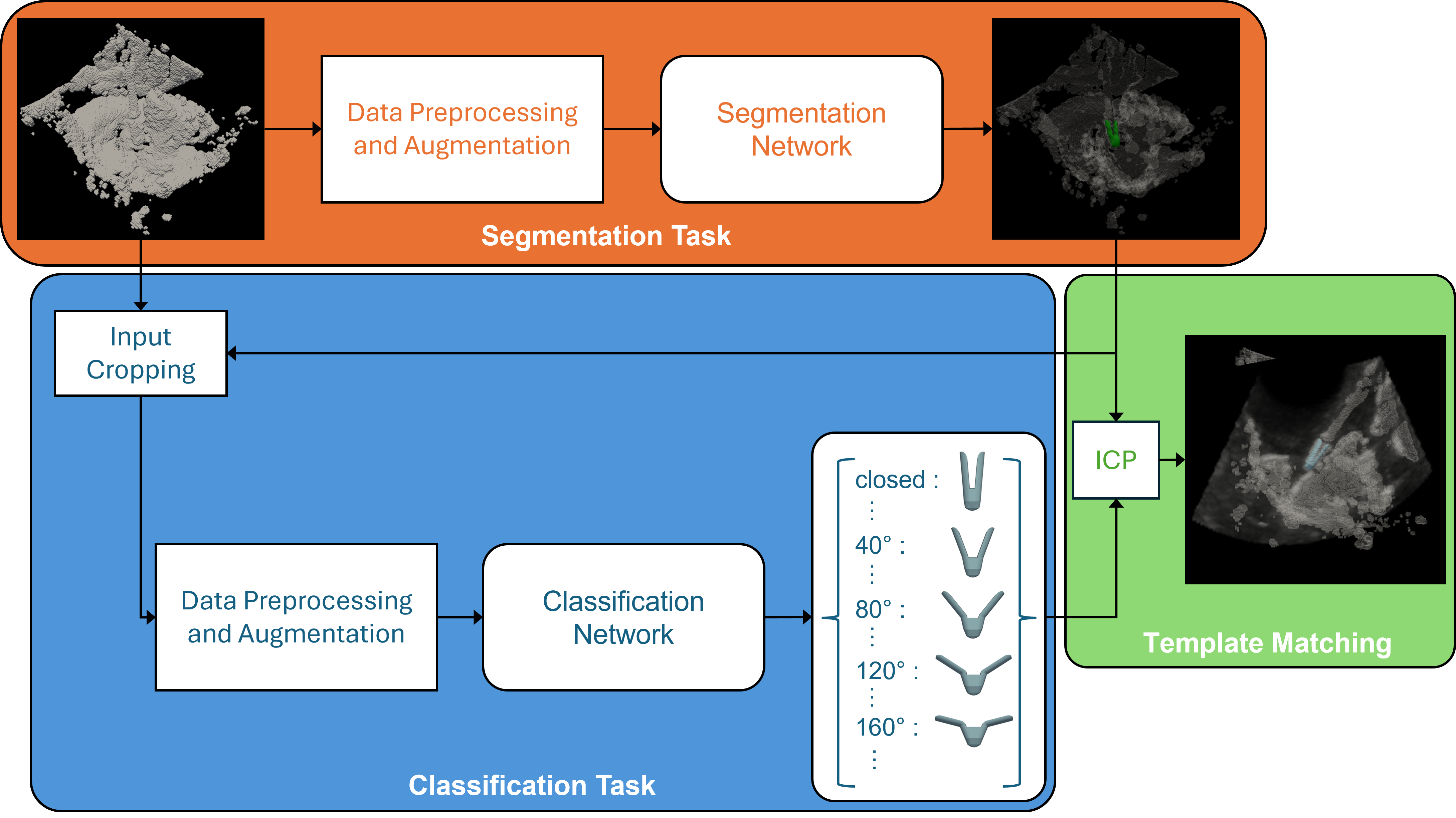}}
    \caption{Schematic representation of the proposed pipeline. Orange box: \textbf{Segmentation Task} — The clip is segmented from 3D \gls{tee} image. Blue Box: \textbf{Classification Task} — The input 3D \gls{tee} image is cropped around the clip using the segmentation output as reference and passed to the classification network to predict the clip configuration. Green Box: \textbf{Template Matching} — A clip template is selected based on the predicted configuration and rigidly registered to the segmentation output. The proposed pipeline enables automatic slicing of the 3D \gls{tee} volume and quantification of the clip status.}
    \label{fig:Pipeline}
\end{figure}

We developed an automated three-stage pipeline for clip detection from 3D \gls{tee} images (Figure \ref{fig:Pipeline}). First, a 3D \gls{cnn} based on a UNet architecture is employed to automatically segment the clip (Figure \ref{fig:Pipeline}, orange box). The largest connected component is extracted from the predicted segmentation to rule out possible spurious unconnected regions. Based on the resulting mask, images are cropped. Second, another 3D \gls{cnn} analyzes cropped images to predict the clip configuration (Figure \ref{fig:Pipeline}, blue box). Third, a template corresponding to the predicted clip configuration is rigidly registered to the output of the 3D segmentation model (Figure \ref{fig:Pipeline}, green box). \\

\subsubsection{Segmentation and Classification Steps}
\label{subsubsec:Segmentation and Classification Steps}

\subsubsection*{(a) Neural Network Architectures}
For the segmentation task, four \gls{cnn} architectures were considered:

\renewcommand\labelitemi{$\vcenter{\hbox{\tiny$\bullet$}}$}

\begin{itemize}
    \item UNet \cite{Ronneberger_2016} with five resolution levels, each of them consisting of a double convolutional block, resulting in approximately 20 milions of trainable parameters.
    \item SegResNet \cite{Myronenko_2019} with asymmetrical design. The encoder was larger, with four down-sampling stages characterized by 1, 2, 2, and 4 ResNet blocks \cite{He_2016}, respectively. These were aimed to enhanced feature extraction. The decoder to reconstruct the segmentation mask was more compact. This architecture resulted in approximately 18 milions of trainable parameters.
    \item Attention UNet \cite{Schlemper_2019} with five resolution levels, each one consisting of a double convolutional block, and attention gates to refine feature maps by focusing on salient regions, enabling the network to prioritize relevant areas for segmentation \cite{Wifstad_2024}. This architecture was characterized by approximately 24 million trainable parameters. 
    \item UNetR \cite{Hatamizadeh_2022} integrating a transformer-based encoder into the UNet framework, featuring five resolution levels and twelve attention heads. This hybrid design allowed the network to capture global, multi-scale information by learning sequential representations of the input volume \cite{Hatamizadeh_2022}. As expected, the UNetR was the heaviest architecture, with the number of trainable parameters exceeding 100 million.\\
\end{itemize}

For the classification task, we utilized ResNet-50 \cite{He_2016} and DenseNet \cite{Huang_2017}, two widely used \gls{cnn} architectures designed for image classification. Both architectures leverage skip connections to improve gradient flow during training, enabling faster convergence and better performance on complex classification tasks. ResNet-50 achieves this through its residual blocks, while DenseNet connects each layer to every subsequent layer within the same block, facilitating feature reuse and efficiency in parameter utilization.

\subsubsection*{(b) Implementation Details}
\label{subsubsec:Implementation Details}

For both the segmentation and classification networks, the dataset was split into training, validation, and testing subsets using a 70:10:20 ratio. Images were resampled to achieve a uniform voxel spacing of 1 mm in all directions. Clip configuration classes were imbalanced within the dataset. To ensure balanced representation, we employed a weighted random sampler during data loading for both segmentation and classification network training, based on configuration frequencies. This approach ensured an even distribution across classes within each batch for both segmentation and classification tasks. In training only, data augmentation was performed on-the-fly, including intensity transforms, i.e., intensity scaling and random Gaussian noise, as well as spatial transforms, i.e., random rotation, random axis flip, and random elastic deformation.\\
During the training of the segmentation network, patches of $128 \times 128 \times 128$ voxels were randomly extracted from the input 3D images. The segmentation networks were trained over 500 epochs with a batch size of 8, using the Adam optimizer, with learning rate initially set to 0.001 and dynamically adjusted using a cosine annealing scheduler. A weighted combination of Dice and Focal losses \citep{Yeung_2022}, with weights of 0.6 and 0.4, respectively, was minimized during the training phase. A 0.1 dropout rate, determined through hyper-parameter search, was selected as it provided the best results in early experiments.\\
Once the segmentation network training had converged, the best-performing segmentation model on the validation set was used to extract homogenous patches of $30 \times 30 \times 30$ voxels from the input images, centered on the segmentation regions. These patches served as input to train the classification networks. The classification models were trained over 600 epochs with batch size of 40, using the Adam optimizer. The initial learning rate was set to 0.00001 and halved every 300 epochs. Cross-entropy loss was used to train the classification networks. \\
The best models on the validation set were saved for inference on the test set for both the segmentation and the classification networks. All models were implemented using PyTorch \cite{Pytorch} and the MONAI library \cite{Cardoso_2022}. Training and inference were performed on an NVIDIA RTX 4090 GPU with 24 GB of memory.

\subsubsection{Template Matching Step}
\label{subsubsec:Template Registration}

Segmentation masks yielded by the first step of the pipeline were refined through a two-step approach.
First, the \gls{cnn}-based segmentation mask of the clip was converted into a 3D surface using the marching cubes algorithm. Second, the surface of the clip template, selected based on the classification performed in the second step of the pipeline, was aligned to the surface of the predicted segmentation mask through  the \gls{icp} algorithm \cite{Zhang_1994}. Specifically, we used the \gls{icp} implementation from the VTK library \cite{vtkBook}. To balance accuracy and computational efficiency, the maximum number of iterations was set to 1,000. 

\subsection{Evaluation}
\label{subsec: evaluation}
The test set consisted of 40 3D \gls{tee} images displaying the clip in several configurations. Clip configurations and number of 3D \gls{tee} images for configuration within the test set are shown in Table \ref{tab: test set ditribution}. To demonstrate the efficacy of the proposed pipeline for clip detection, we separately evaluated the segmentation and classification performances. For the segmentation task, the predicted segmentations were compared with the refined clip \glspl{gt} described in Section \ref{subsubsec: Clip Refined Ground Truth}. The largest connected component was extracted from the predicted segmentations before computing the evaluation metrics. Standard segmentation metrics were computed, including the Dice score, \gls{asd}, and \gls{hdf95}. For the classification task, we evaluated the performance by computing precision, recall, and F1-score for each class in the test set, as well as the weighted average performance across all classes defined by \[
\text{Weighted Average} = \sum_{i=1}^N \frac{n_i}{N_{\text{total}}} \cdot M_i
\]
where $n_{i}$ is the number of true instances for class $i$, $M_{i}$ is the metric value (e.g., precision, recall, or F1-score) for class $i$, and $N_{\text{total}} = \sum_{i=1}^N n_i$ is the total number of instances across all classes. Using the best classification architecture between ResNet-50 and DenseNet, we recomputed the segmentation performance after registering the predicted template based on the classification network's output.

\begin{table}[H]
\centering
\resizebox{\textwidth}{!}{
\begin{tabular}{lclllllllll}
\hline
& \multicolumn{10}{c}{Clip's Configurations in the Test Set}\\
\multicolumn{1}{c}{Angle Configuration} & \multicolumn{1}{l}{\textbf{0°}} & \textbf{20°}          & \textbf{40°}          & \textbf{60°}          & \textbf{80°}          & \textbf{100°}         & \textbf{120°}         & \textbf{140°}         & \textbf{160°}         & \textbf{180°}         \\ \hline
Number of Instances                     & 19                              & \multicolumn{1}{c}{0} & \multicolumn{1}{c}{1} & \multicolumn{1}{c}{2} & \multicolumn{1}{c}{3} & \multicolumn{1}{c}{4} & \multicolumn{1}{c}{2} & \multicolumn{1}{c}{1} & \multicolumn{1}{c}{1} & \multicolumn{1}{c}{5} \\ \hline
\end{tabular}
}
\caption{Distribution of clip configurations within the test set.}
\label{tab: test set ditribution}
\end{table}

\section{Results}
\label{sec:results}

\subsection{Segmentation Performance}
\label{subsec: Segmentation Performance}
\begin{figure}[ht]
    \centering
    \centerline{\includegraphics[width=\textwidth]{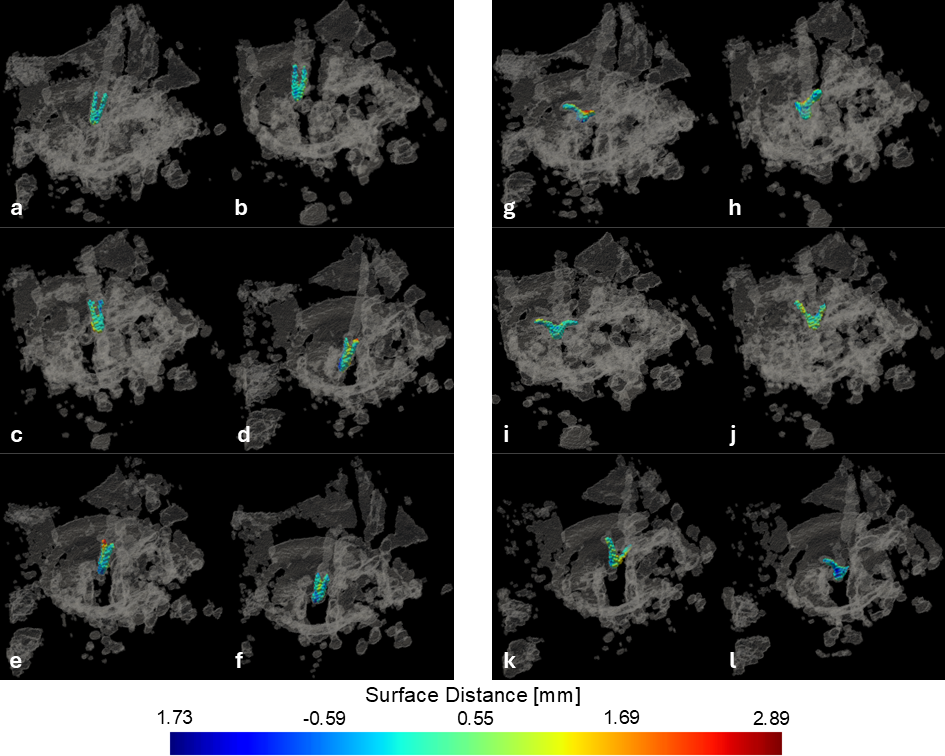}}
    \caption{Segmentation examples provided by the Attention UNet for closed (a-f) and open clip configurations (g-l). Surface distance heatmaps computed respect to the GT, are overlaid on the segmentation masks.}
    \label{fig:Segmentation Performance}
\end{figure}

Table \ref{tab: Segmentation Performance} summarizes the average performance across the test set achieved by the network architectures considered for clip segmentation. Among all the models, Attention UNet demonstrated the best performance achieving a Dice score of $0.62 \pm 0.14$, \gls{asd} of $0.76 \pm 1.03$ mm and \gls{hdf95} $2.44 \pm 2.40$ mm. The simple UNet achieved comparable performance but exhibited a lower Dice score and higher distance metrics compared to Attention UNet. In contrast, SegResNet and UNetR showed weaker results, with Dice score below $0.6$, \gls{asd} around 1 mm, and \gls{hdf95} values close to 4 mm.\\
Figure \ref{fig:Segmentation Performance} provides examples of clip segmentations obtained using the Attention UNet. It depicts the device in a closed configuration (Figure \ref{fig:Segmentation Performance}, a-f) and in several open configurations (Figure \ref{fig:Segmentation Performance}, g-l). Overall, the segmentation network successfully localized the clip and captured its shape. The bottom part of the clip was the least challenging to detect, as indicated by lower distance errors vs. \glspl{gt}. Conversely, the upper part proved more challenging, yielding higher distance errors, especially in open configurations where one of the clip’s arms was occasionally missed (Figure \ref{fig:Segmentation Performance}, g, h and l).

\begin{table}[H]
\centering
\resizebox{\textwidth}{!}{
\begin{tabular}{lcccc}
\hline
 Metric      & UNet        & Attention UNet   & SegResNet   & UNetR       \\
\hline
 Dice score  & 0.59 ± 0.16 & \textbf{0.62 ± 0.14}      & 0.53 ± 0.12 & 0.53 ± 0.19 \\
 ASD (mm)    & 0.81 ± 1.00 & \textbf{0.76 ± 1.03}      & 1.04 ± 1.10 & 1.32 ± 3.93 \\
 95\% HD (mm) & 2.73 ± 2.88 & \textbf{2.44 ± 2.40}      & 3.88 ± 2.91 & 4.35 ± 5.18 \\
\hline
\end{tabular}
}
\caption{Evaluation performance across the test set. \Gls{asd} and \gls{hdf95} are reported for the UNet, AttentionUNet, SegResNet and UNetR. The best value for each segmentation metric across all models is highlighted in bold.}
\label{tab: Segmentation Performance}
\end{table}

\subsection{Classification Performance}
\label{subsec: Classification Performance}

\begin{figure}[H]
    \centering
    \centerline{\includegraphics[width=\textwidth]{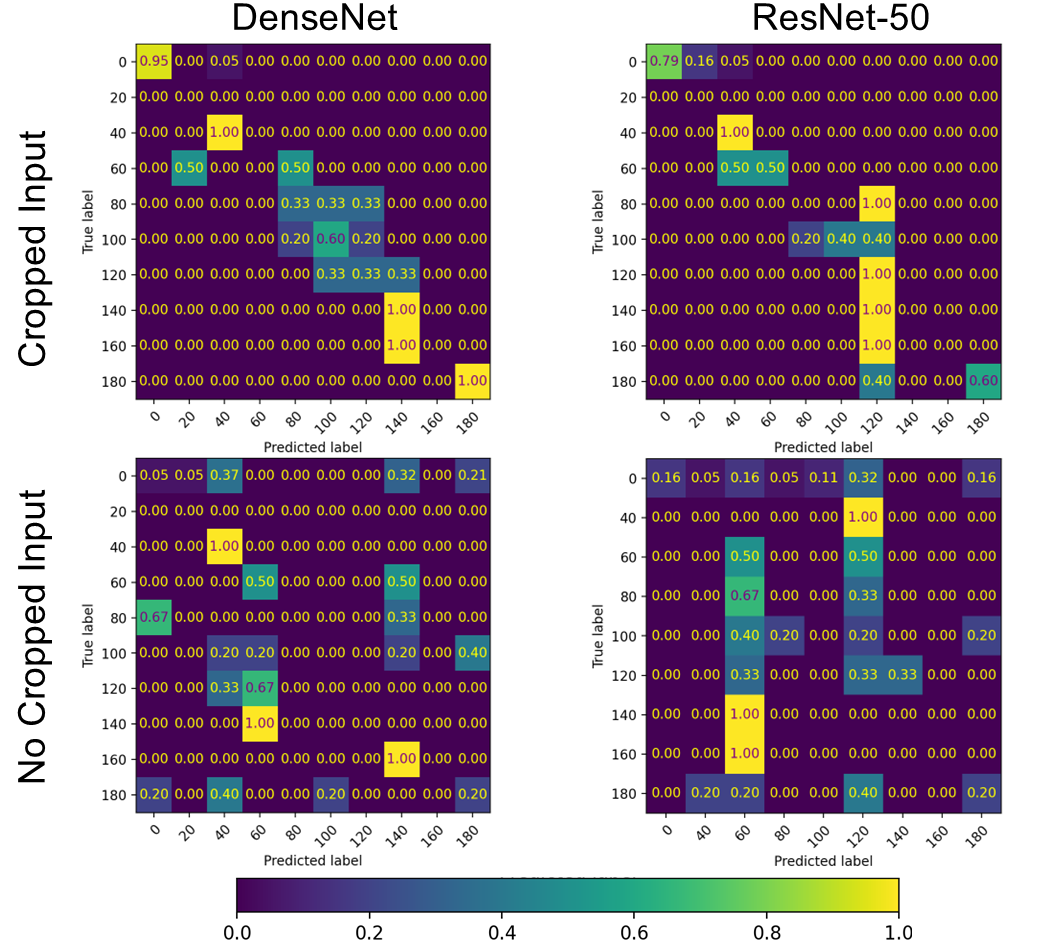}}
    \caption{Confusion matrices for DenseNet and ResNet-50 classification performance across the test set. Matrices are shown for both cropped and uncropped input data. Normilized predictions are shown in each matrices.}
    \label{fig:Confusion Matrix}
\end{figure}

Table \ref{tab: Classification performance} presents the classification performance achieved by DenseNet and ResNet-50 across the test set for cropped and uncropped input data according to the segmentation output. For this evaluation, the Attention UNet was used as the segmentation network, as it achieved the best performance in the segmentation task. 
Both DenseNet and ResNet-50 performed significantly better with cropped input data, achieving a weighted average F1-score of 0.74 and 0.66, respectively, across the different clip configurations. DenseNet outperformed ResNet-50 in precision (0.75 vs. 0.80) but had slightly lower recall (0.75 vs. 0.63). For the 0° and 180° configurations, DenseNet achieved optimal precision, recall, and F1-scores (F1 = 0.97 and 1.0, respectively). Worse results were obtained for 80° and 120° configurations, achieving F1-scores of 0.33. The model showed poor classification performance for limited-support classes such as 60° and 160°. ResNet-50 excelled in classifying 40° (F1=0.5) and 60° (F1=0.67) configurations, but underperformed in configurations with limited supports, such as 140° and 160°.\\
For uncropped input data, both DenseNet and ResNet-50 struggled significantly, achieving a weighted average F1-score of 0.08 and 0.17. respectively. The models exhibited near-zero precision and recall across most configurations, failing to accurately classify configurations such as 0°, 60°, and 100°.\\
Figure \ref{fig:Confusion Matrix} shows the confusion matrices for the classification performances achieved by DenseNet and ResNet-50 across the test set for cropped and uncropped input data. The confusion matrices confirm the previous observations, showing significantly better performance with cropped input data for both DenseNet and ResNet-50. The matrices for cropped input are more diagonally dominant, reflecting a higher rate of correct predictions. DenseNet's confusion matrix for cropped input shows less off-diagonal entries than the confusion matrix of ResNet-50, indicating more consistent classification. Misclassifications primarily resulted in errors of one configuration step (20°), particularly for classes such as 80°, 100°, and 160°. Larger errors of up to 40° were observed for classes such as 0°, 60°, and 120°. 
In contrast, with uncropped input, the confusion matrices for both models display substantial off-diagonal entries for both DenseNet and ResNet-50, reflecting significant misclassifications. Predictions are often biased toward dominant configurations, such as 0° and 180°, and errors are distributed across the matrix.

\begin{table}[H]
\centering
\resizebox{\textwidth}{!}{
\begin{tabular}{lcccccccccccccc}
\hline
                 & \multicolumn{6}{c}{Cropped Input}                                                                                                                                                     & \multicolumn{1}{l}{} & \multicolumn{6}{c}{Uncropped Input}                                                                                                                                                  & \multicolumn{1}{l}{}        \\ \cline{2-7} \cline{9-14}
                 & \multicolumn{3}{c}{DenseNet}                                                              & \multicolumn{3}{c}{ResNet-50}                                                             & \multicolumn{1}{l}{} & \multicolumn{3}{c}{DenseNet}                                                              & \multicolumn{3}{c}{ResNet-50}                                                             & \multicolumn{1}{l}{}        \\ \hline
Configuration    & \multicolumn{1}{l}{Precision} & \multicolumn{1}{l}{Recall} & \multicolumn{1}{l}{F1-score} & \multicolumn{1}{l}{Precision} & \multicolumn{1}{l}{Recall} & \multicolumn{1}{l}{F1-score} & \multicolumn{1}{l}{} & \multicolumn{1}{l}{Precision} & \multicolumn{1}{l}{Recall} & \multicolumn{1}{l}{F1-score} & \multicolumn{1}{l}{Precision} & \multicolumn{1}{l}{Recall} & \multicolumn{1}{l}{F1-score} & \multicolumn{1}{l}{Support} \\ \hline
0°               & 1.0                           & 0.95                       & 0.97                         & 1.0                           & 0.78                       & 0.88                         &                      & 0.25                          & 0.05                       & 0.09                         & 0.0                           & 0.0                        & 0.0                          & 19                          \\
20°*              & 0.0                           & 0.0                        & 0.0                          & 0.0                           & 0.0                        & 0.0                          &                      & 0.0                           & 0.0                        & 0.0                          & 0.0                           & 0.0                        & 0.0                          & 0                           \\
40°              & 0.5                           & 1.0                        & 0.67                         & 0.33                          & 1.0                        & 0.5                          &                      & 0.08                          & 1.0                        & 0.15                         & 0.0                           & 0.0                        & 0.0                          & 1                           \\
60°              & 0.0                           & 0.0                        & 0.0                          & 1.0                           & 0.5                        & 0.67                         &                      & 0.2                           & 0.5                        & 0.28                         & 0.0                           & 0.0                        & 0.0                          & 2                           \\
80°              & 0.33                          & 0.33                       & 0.33                         & 0.0                           & 0.0                        & 0.0                          &                      & 0.0                           & 0.0                        & 0.0                          & 0.03                          & 0.03                       & 0.01                         & 3                           \\
100°             & 0.6                           & 0.6                        & 0.6                          & 1.0                           & 0.4                        & 0.57                         &                      & 0.0                           & 0.0                        & 0.0                          & 0.0                           & 0.0                        & 0.0                          & 5                           \\
120°             & 0.33                          & 0.33                       & 0.33                         & 0.25                          & 1.0                        & 0.4                          &                      & 0.0                           & 0.0                        & 0.0                          & 0.0                           & 0.0                        & 0.0                          & 3                           \\
140°             & 0.33                          & 1                          & 0.5                          & 0.0                           & 0.0                        & 0.0                          &                      & 0.0                           & 0.0                        & 0.0                          & 0.0                           & 0.0                        & 0.0                          & 1                           \\
160°             & 0.0                           & 0.0                        & 0.0                          & 0.0                           & 0.0                        & 0.0                          &                      & 0.0                           & 0.0                        & 0.0                          & 0.0                           & 0.0                        & 0.0                          & 1                           \\
180°             & 1.0                           & 1.0                        & 1.0                          & 1.0                           & 0.6                        & 0.75                         &                      & 0.14                          & 0.2                        & 0.17                         & 0.0                           & 0.0                        & 0.0                          & 5                           \\ 
Weighted Average & 0.75 & 0.75 & 0.74 & 0.80 & 0.63 & 0.66 & &  0.15 & 0.1 & 0.08 & 0.51 & 0.15 & 0.17 & 40                          \\ \hline
\end{tabular}
}
\caption{Classification performance of DenseNet and ResNet-50 for clip configurations across the test set. Results are reported for cropped and uncropped input data. Metrics include precision, recall, and F1-score for each configuration, along with the corresponding weighted averages. The cropped input uses segmentation-assisted preprocessing, while the uncropped input relies on randomly extracted patches. *Classification performance for 20° configuration is consistently zero because this class is absent from the test set.}
\label{tab: Classification performance}
\end{table}

\subsection{Segmentation Performance after Refinement}
\label{subsec: Classification Performance after Refinement}

\begin{figure}[H]
    \centering
    \centerline{\includegraphics[width=\textwidth]{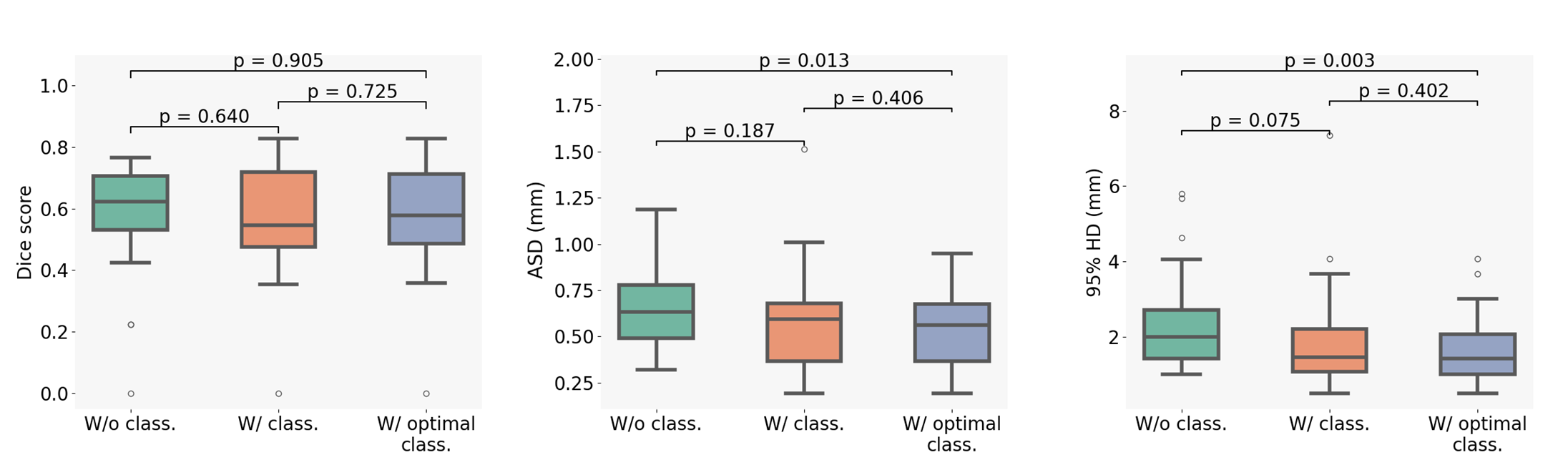}}
    \caption{Boxplots for segmentation performance achieved by the Attention UNet with and without segmentation refinement through template matching. For Dice score, \gls{asd} and \gls{hdf95} the boxplots compare performance metric for the raw segmentation output (W/o classification, green box), the refined segmentation output based on the predicted configuration from DenseNet (W/ classification, orange box) and the the refined segmentation based on the actual configuration (W/ optimal classification, purple box). Level of significance is reported above the box plot respect to the performance metric for the raw segmentation output.}
    \label{fig:Boxplots}
\end{figure}

\begin{figure}[H]
    \centering
    \centerline{\includegraphics[width=\textwidth]{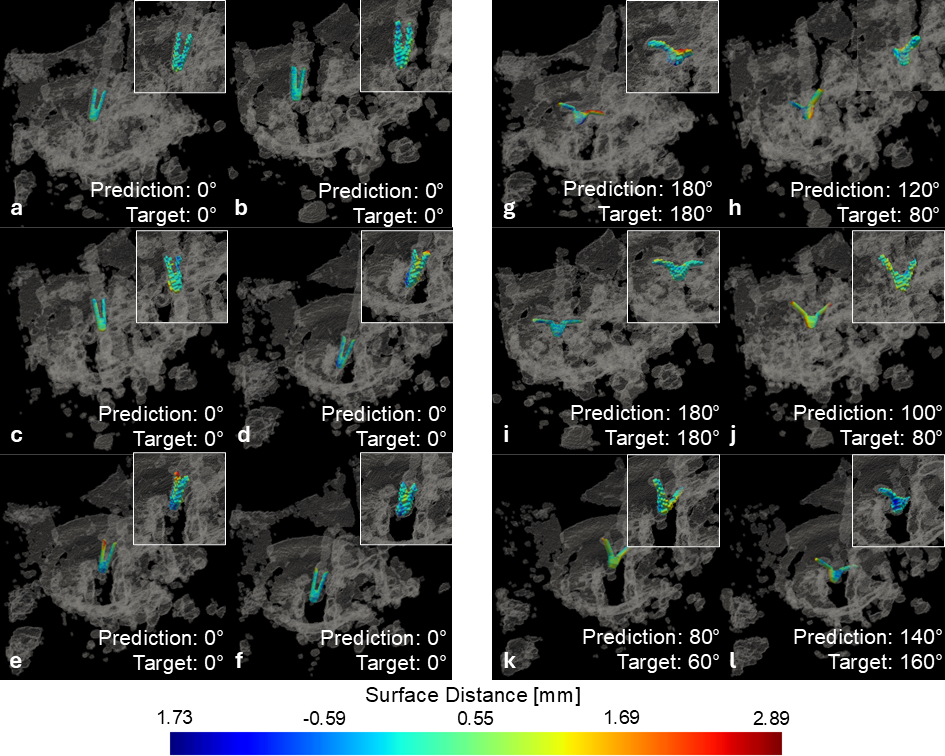}}
    \caption{Segmentation examples provided by Attention UNet for closed configurations (a-f) and open clip configurations (g-l). Each case includes a zoomed view of the raw segmentation and reports the predicted configuration from DenseNet alongside the target configuration used for template selection. Surface distance heatmaps, computed relative to the \gls{gt}, are overlaid on the segmentation masks.}
    \label{fig:Segmentation Performance After}
\end{figure}

Table \ref{tab: Refined Segmentation perfromance} summarizes the average performance across the test set achieved by the network architectures after template registration through the \gls{icp} algorithm. Templates were selected based on the classification outputs provided by the DenseNet, which achieved the best performance in the classification task. For this step, input data for the DenseNet were cropped according to the segmentation output provided by the evaluated segmentation network.\\ Similar to the raw segmentation performance reported previously, Attention UNet demonstrated the best performance achieving a Dice score of $0.59 \pm 0.16$, \gls{asd} of $0.75 \pm 1.14$ mm and \gls{hdf95} $2.05 \pm 2.50$ mm. 
UNet achieved comparable distance metrics, with an \gls{asd} of $0.75\pm1.20$ mm and a \gls{hdf95} of $2.07\pm2.33$ mm, but showed a slightly lower average Dice score of $0.57\pm0.17$ compared to Attention UNet. In contrast, SegResNet and UNetR slightly underperformed as compared to the other architectures, achieving a Dice score slightly above 0.5, on average. UNetR showed the weakest overall performance, with an average \gls{asd} value and the average \gls{hdf95} value above 1 mm and 3 mm, respectively. \\ 
Compared to the raw segmentation results (Table \ref{tab: Segmentation Performance}), template registration using the \gls{icp} algorithm notably improved \gls{asd} and \gls{hdf95} values. For example, SegResNet showed the most notable improvement in \gls{asd}, which decreased from 
$1.04 \pm 1.10$ mm to $0.90 \pm 1.09$ mm, and in 95\% HD, which decreased from $3.88 \pm 2.91$ mm to $2.49 \pm 2.08$ mm. Similarly, UNetR exhibited a significant reduction in \gls{hdf95}, improving from $4.35\pm5.18$ mm to $3.11\pm5.07$ mm. However, the Dice score for all segmentation networks did not improve; rather, it slightly worsened after refinement through template registration.\\
Figure \ref{fig:Boxplots} compares the segmentation performance of Attention UNet across three scenarios: the raw segmentation output, the refined segmentation output based on the predicted classification provided by DenseNet, and the refined segmentation output based on the actual classification. The Dice score slightly worsened after segmentation refinement, with the median value decreasing from 0.62 to 0.54. In contrast, \gls{asd} and \gls{hdf95} improved, with median values decreasing from 0.63 mm to 0.59 mm and from 2.0 mm to 1.45 mm, respectively. Segmentation performance based on the actual classification yielded median values of 0.57, 0.56 mm, and 1.41 mm, respectively, which were comparable to those obtained using the predicted classification.\\
Figure \ref{fig:Segmentation Performance After} provides qualitative examples of clip segmentations obtained with Attention UNet after refinement through template matching. It depicts the device in a closed configuration (Figure \ref{fig:Segmentation Performance After}, a-f) and in several open configurations (Figure \ref{fig:Segmentation Performance After}, g-l). Each case shows a zoomed view of the raw segmentation and reports the predicted configuration from DenseNet alongside the target configuration used for template selection. Overall, template matching improved segmentation outputs by enhancing consistency with \glspl{gt}, as reflected in improved distance metrics. Template matching successfully restored small details, such as the clip arms, that were sometimes lost in raw segmentation outputs, particularly in open configurations (Figure \ref{fig:Segmentation Performance After}, g and l). Incorrect classification predictions, and consequently incorrect template matching, led to slight worsening of distance metrics (Figure \ref{fig:Segmentation Performance After}, h, j, k and l). However, these errors were negligible, as they involved small differences in clip configurations. These variations are indiscernible given the resolution and image quality of echocardiography and are not significant in the context of intraprocedural guidance.

\begin{table}[H]
\centering
\resizebox{\textwidth}{!}{
\begin{tabular}{lcccc}
\hline
 Metric      & UNet        & Attention UNet   & SegResNet   & UNetR       \\
\hline
 Dice score  & 0.57 ± 0.17 & \textbf{0.59 ± 0.16}      & 0.51 ± 0.17 & 0.52 ± 0.19 \\
 ASD (mm)    & 0.75 ± 1.20 & \textbf{0.75 ± 1.14}      & 0.90 ± 1.09 & 1.38 ± 3.92 \\
 95\% HD (mm) & 2.07 ± 2.33 & \textbf{2.05 ± 2.50}      & 2.49 ± 2.08 & 3.11 ± 5.07 \\
\hline
\end{tabular}
}
\caption{Evaluation performance across the test set after segmentation refinement through template matching. Template selection was based on predictions from DenseNet. Average Dice score, \gls{asd} and \gls{hdf95} are reported for the UNet, AttentionUNet, SegResNet and UNetR. The best value for each segmentation metric across all models is highlighted in bold.}
\label{tab: Refined Segmentation perfromance}
\end{table}

\subsection{Inference Time}
\label{Inference Time}

The proposed pipeline takes $2.55\pm0.58$ s to process a 3D \gls{tee} volume and detect the clip within it. The required time is subdivided as follows: $0.01 \pm 0.01$ s for the segmentation task, $0.09 \pm 0.03$ s for the classification task and $2.44\pm0.58$ s for model refinement through template matching. The notable time-expense of the final step results from leveraging an iterative algorithm, i.e., \gls{icp}: its time-expense can increase massively depending on the number of points considered for surface alignment, the convergence threshold, and the maximum number of iterations allowed for convergence.

\section{Discussion}
\label{sec: Discussion}

We presented an automated pipeline for clip detection from intraoperative 3D \gls{tee} images, enabling accurate localization of the device, configuration quantification, and enhanced 3D visualization within the intracardiac space. To the best of our knowledge, this is the first attempt to automate detection of a device of such complexity from intraoperative 3D \gls{tee} images. The pipeline offers two significant advancements for MitraClip procedures. First, it provides a fast detection of the clip, enabling localization in the intracardiac space for the operator. Second, it provides a fast classification of the clip pose, hence essential information for device orientation and position. Of note, this information could be combined with the result of automated \gls{mv} segmentation and reconstruction to yield quantitative information on the pose of the clip relative to the target region.\\
We trained and evaluated four different 3D \gls{cnn} architectures for clip segmentation: UNet, Attention UNet, SegResNet, and UNetR. While all these models share a common backbone structure based on the traditional UNet’s encoding-decoding architecture, each of them incorporates distinctive features tailored to address specific challenges in medical imaging, including echocardiography. These architectures are well-established for medical image segmentation. Traditional UNet \cite{Jia_2022} and its custom variants \cite{Yang_2019, Yang_2019b, Yang_2019c, Yang_2020, Yang_2021} have been widely applied to catheter segmentation in echocardiography, while Attention UNet \cite{Wifstad_2024} and SegResNet \cite{Munafo_2025} have been successfully used for \gls{mv} segmentation from 2D and 3D \gls{tee}. UNetR incorporates multi-head attention mechanisms typical of Transformer architectures \cite{Vaswani_2017}, which have shown promising performance in segmentation tasks beyond their original use in natural language processing. These networks were trained using manually generated \gls{gt} segmentations, refined by matching with \gls{cad}-based template models of a real MitraClip device. This refinement step ensured that the \gls{gt} segmentations preserved the actual dimensions of the device, enhancing reliability and precision in detection. This approach was required because accurate manual delineation of implantable devices is particularly challenging in \gls{tee} images due to noise, artifacts, and the presence of metallic catheter components, which often lead to over-segmentation and false positives that can compromise localization accuracy \cite{Mastmeyer_2017}. Among the evaluated architectures, Attention UNet demonstrated the best segmentation performance, on average. As shown in Figure \ref{fig:Segmentation Performance}, Attention UNet reliably segmented the clip, accurately reconstructing the clip's tip and arms with minimal error vs. \glspl{gt} in closed configuration (Figure \ref{fig:Segmentation Performance} a-f). However, in open configurations, the network occasionally failed to fully reconstruct the clip, missing one of the two arms. This typically occurred when the clip was positioned near heart structures, where the limited image contrast hindered clear clip visualization (Figure \ref{fig:Segmentation Performance} g and l). Attention UNet benefited from its attention gates, which enabled better localization of salient features \cite{Schlemper_2019}, improving segmentation accuracy for challenging tasks like clip detection, where the device occupies a small region within the broader field of view in 3D \gls{tee}. This architecture guaranteed state-of-the-art performance for cardiac catheter detection from 3D \gls{tee}, comparable to previous studies \cite{Jia_2022, Yang_2019b, Yang_2019c, Yang_2020, Yang_2021} that employed UNet-based architectures and reported an average Dice score ranging between 0.57 and 0.7, and average HD between 7 and 1 voxels. Unlike those studies, our approach did not rely on additional preprocessing nor on architecture enhancements, such as multi-planar slicing \cite{Yang_2019}, patch-of-interest extraction \cite{Yang_2019c}, contextual information enhancement \cite{Yang_2019b, Yang_2020}, or spatial complexity reduction using projection layers \cite{Yang_2021} in UNet's decoder path. Furthermore, previous studies focused on segmenting simpler structures, such as ablation catheters \cite{Jia_2022} or guide wires \cite{Yang_2019b, Yang_2019c, Yang_2020, Yang_2021}, which typically maintain a straight configuration during procedures. In contrast, the MitraClip system includes a highly steerable delivery catheter and a clip with dynamically adjustable arms. Hence, is presents unique challenges requiring more sophisticated detection and segmentation strategies.\\
For the classification task, we compared the performance of DenseNet and ResNet-50. Both architectures benefited from segmentation-assisted input cropping, which allowed the classifier to focus on features extracted from the region of interest provided by the segmentation network. This approach improved the classification accuracy without requiring additional operations, as the segmentation network was already part of the pipeline. Using cropped input data, DenseNet outperformed ResNet-50, achieving a higher weighted average F1-score (0.75 vs. 0.63). The dense connectivity of DenseNet likely contributed to more robust feature reuse, resulting in superior classification accuracy. Optimal precision, recall, and F1-scores were achieved for some configurations, particularly closed and 180°. However, DenseNet struggled to classify configurations with limited support, such as 60° and 160°. Despite these challenges, DenseNet’s errors were generally minor, with most misclassifications differing by only 20° and a maximum error of 40° (Figure \ref{fig:Confusion Matrix}). These small misclassification errors are unlikely to impact clinical outcomes significantly, as they do not hinder the overall understanding of the clip configuration.\\
In the final step of our proposed pipeline, template registration was employed based on the predicted configuration to enhance segmentation accuracy. For simpler catheter shapes, such as ablation catheters or guide wires that maintain a mostly straight configuration, model fitting was previously proposed to improve segmentation output \cite{Yang_2019}. However, the higher geometrical complexity of the clip and its broad range of configurations required template matching instead. This approach proved effective in improving distance metrics. For instance, SegResNet's \gls{asd} decreased from $1.04 \pm 1.10$ to $0.90\pm1.09$ mm and the \gls{hdf95} decreased from $3.88 \pm 2.91$ to $2.49\pm2.08$ mm. Similarly, UNetR exhibited a reduction in \gls{hdf95} from $4.35\pm5.18$ mm to $3.11\pm5.07$ mm. Unexpectedly, for all architectures the average Dice score slightly worsened after refinement. However, as shown for Attention UNet, this decrease was not statistically significant ($p>0.05$, Figure \ref{fig:Boxplots}). Conversely, improvements in distance metrics, which better reflect structural alignment, were more pronounced. Although not all differences were statistically significant, this trend suggests that the limited dataset size may have influenced the analysis. Notably, Attention UNet achieved median \gls{asd} values of 0.59 mm and 0.56 mm, and median \gls{hdf95} values of 1.45 mm and 1.41 mm, respectively, when using the predicted and actual classifications. These findings strongly suggest that the observed misclassifications of clip configuration negligibly contributed to distance errors and did not significantly affect the accuracy of clip reconstruction vs. \glspl{gt}.  
Overall, template matching was effective in restoring structural details and original shape of the detected device, preserving key landmarks, such as clip tip and arms, to be detected in 3D \gls{tee} for intraprocedural guidance. This behavior explains the observed improvement in distance metrics, which are more sensitive to shape variations than overlap-based metrics like the Dice score. For closed configurations (Figure \ref{fig:Segmentation Performance After}, a-f), template matching accurately reconstructed the clip’s arms, often missed in raw segmentations due to noise or low contrast. In open configurations, template matching improved alignment by restoring the original shape in cases where one arm was missed, enhancing agreement with \glspl{gt} (Figure \ref{fig:Segmentation Performance After}, g and l). However, as template matching relies on the \gls{icp} algorithm, its effectiveness depends on the reliability of the target structure (in this case, the predicted segmentation) and the topological similarity between the template and the target structure. Incomplete segmentations or erroneous classifications can lead to suboptimal registration, potentially explaining the lack of Dice score improvement after segmentation refinement. Unlike distance metrics, the Dice score heavily relies on the degree of overlap between the predicted and target structures, making it less sensitive to cases where template matching restores shape details but fails to achieve optimal alignment with the \glspl{gt}. Despite these subtleties, the pipeline would still enable automatic extraction of optimal 2D views to visualize the clip, since these could be positioned and oriented in the 3D \gls{tee} volume based on the pose of the reconstructed clip. Errors associated with template selection were minor and did not compromise the accurate interpretation of the clip configuration.\\ 
The proposed pipeline has the potential to provide accurate and reliable guidance during MitraClip procedures in near-real-time ($2.55\pm0.58$ seconds). 
It enhances image acquisition and interpretation in intracardiac scenarios, which are typically hindered by echocardiographic limitations such as noise and low contrast. 
Moreover, our automated pipeline could be extended to other devices for \gls{teer}, e.g., the Pascal Precision System \cite{Garcia_2024} by Edwards Lifesciences, or for transcatheter tricuspid valve repair, e.g., Triclip \cite{Sorajja_2023} by Abbott, with minimal modifications. This flexibility broadens its applicability and offers significant potential to improve procedural outcomes across various interventional cardiology applications.

\subsection{Limitations and Future Works}
\label{sec: Limitations and Future Works}

This study utilized a dataset acquired \textit{in vitro} on a heart simulator, which provided a controlled and reliable environment for evaluating the proposed pipeline. However, to comprehensively assess the performance of the method in real-world scenarios, future research should focus on in-vivo data acquired from human subjects. Such data will enable the evaluation of the pipeline's robustness and generalization capability under clinical conditions, including variations in patient anatomy and imaging quality, and the assessment of its impact on the current clinical workflow. If successfully translated, the proposed pipeline has the potential to enhance procedural effectiveness, reduce procedural times, and mitigate the risk of adverse outcomes by providing real-time quantitative feedback to operators. \\
Another limitation lies in the dataset's imbalanced representation of clip configurations, which posed challenges in model training and evaluation. To mitigate this, we applied oversampling during neural networks training. While this approach reduced the impact of imbalance, it did not entirely eliminate errors, particularly for intermediate configurations. Although these errors were not critical for the final application, they indicate room for improvement. Future work could focus on acquiring a larger and more diverse dataset with a balanced distribution of clip configurations to enhance the model's robustness. Additionally, the model exhibited better performance in classifying fully closed and fully open configurations, likely due to their higher representation in the dataset and their more distinguishable features in 3D \gls{tee} volumes. \\
The segmentation and classification steps of the pipeline achieved fast inference times suitable for intraprocedural guidance. However, the segmentation refinement step, which relies on the \gls{icp} algorithm, significantly increased computational time. The \gls{icp} algorithm's computational intensity and dependence on the quality of segmentation output present challenges; imprecise segmentations can lead to suboptimal refinement. To address these challenges, a potential future direction could involve replacing the segmentation network with a model that directly infers the orientation and configuration of the device. End-to-end approaches for predicting 3D object orientation are well-established in general computer vision \cite{Qi_2017} and could be adapted for this application. Such an approach could eliminate the need for segmentation refinement, thereby streamlining the process. 

\section{Conclusion}
\label{sec:conclusion}
We presented an automated pipeline for accurate clip detection and pose classification from intraprocedural 3D \gls{tee} images. The proposed method demonstrated high precision in device localization and reliable configuration assessment, addressing critical challenges in the real-time interpretation of echocardiographic data during MitraClip procedures. By simplifying and automating these tasks, this method has the potential to enhance procedural accuracy and efficiency, reducing the cognitive and physical workload on operators. Improved localization and configuration assessment directly support critical clinical decisions, potentially minimizing procedural times and mitigating the risks associated with adverse outcomes.

\section*{Acknowledgements}
\label{sec:acknowledgements}
This work was supported by the European Union’s Horizon 2020 research and innovation program, under the project ARTERY, grant agreement No. 101017140.



\newpage
\bibliographystyle{ieeetr} 
\bibliography{references}

\end{document}